\documentclass[]{PoS}

\title{Simulating Cherenkov Telescope Array observation of RX J1713.7--3946}

\ShortTitle{CTA simulations of RX~J1713.7--3946}

\author{
\speaker{T. Nakamori}$^{a}$,
H.~Katagiri $^b$,
H.~Sano $^c$,
R.~Yamazaki $^d$,
Y.~Ohira $^d$,
A.~Bamba $^d$,
Y.~Fukui $^c$,
K.~Mori  $^e$,
S.-H.~Lee $^f$,
Y.~Fujita $^g$,
H.~Tajima $^h$,
T.~Inoue $^i$,
S.~Gunji $^a$,
Y.~Hanabata $^j$,
M.~Hayashida $^j$,
H.~Kubo $^k$,
J.~Kushida $^l$,
S.~Inoue $^{m}$,
K.~Ioka $^{n}$,
K.~Kohri $^{n}$,
K.~Murase $^{o}$,
S.~Nagataki $^{m}$,
T.~Naito $^{p}$,
A.~Okumura $^{h,q,r}$,
T.~Saito $^{k}$,
M.~Sawada $^{d}$,
T.~Tanaka $^{k}$,
Y.~Terada $^{s}$,
Y.~Uchiyama $^{t}$,
S.~Yanagita $^{b}$,
T.~Yoshida $^{b}$,
T.~Yoshikoshi $^{j}$,
for the CTA Consortium\footnote{Full consortium author list at http://cta-observatory.org}\\
        E-mail: \email{nakamori@sci.kj.yamagata-u.ac.jp}

{\footnotesize
$^{a}$ Yamagata University, 1-4-12 Kojirakawa, Yamagata 990-8560, Japan;
$^{b}$ Ibaraki University, 2-1-1 Bunkyo, Mito, Ibaraki 310-8512, Japan;
$^{c}$ Nagoya University, Furo-cho, Chikusa-ku, Nagoya, Aichi 464-8602, Japan;
$^{d}$ Aoyama Gakuin University, Sagamihara, Kanagawa, 252-5258, Japan;
$^{e}$ University of Miyazaki, 1-1 Gakuen Kibana-dai Nishi, Miyazaki 889-2192, Japan;
$^{f}$ ISAS/JAXA, 3-1-1 Yoshinodai, Chuo-ku, Sagamihara, Kanagawa 252-5210, Japan;
$^{g}$ Osaka University, Toyonaka, Osaka 560-0043, Japan;
$^{h}$ STE Laboratory, Nagoya University, Furo-cho, Chikusa-ku, Nagoya 464-8601, Japan;
$^{i}$ NAOJ, 2-21-1 Osawa, Mitaka, Tokyo, 181-8588, Japan;
$^{j}$ ICRR, University of Tokyo, 5-1-5 Kashiwanoha, Kashiwa, Chiba 277-8582, Japan;
$^{k}$ Kyoto University, Kitashirakawa-Oiwake, Sakyo-ku, Kyoto 606-8502, Japan;
$^{l}$ Tokai University, 4-1-1 Kita-Kaname, Hiratsuka, Kanagawa 259-1292, Japan;
$^{m}$ RIKEN, 2-1 Hirosawa, Wako, Saitama 351-0198, Japan;
$^{n}$ KEK, 1-1 Oho, Tsukuba 305-0801, Japan;
$^{o}$ Institute for Advanced Study, 1 Einstein Dr., Princeton, NJ 08540, USA;
$^{p}$ Yamanashi Gakuin University, Sakaori 2-4-5, Kofu-shi, Yamanashi 400-8575, Japan;
$^{q}$ University of Leicester, University Road, Leicester, LEI 7RH, UK;
$^{r}$Max-Planck-Institut f\"{u}r Kernphysik, P.O.Box 103980, D 69029 Heidelberg, Germany;
$^{s}$ Saitama University, 255 Simo-Ohkubo, Sakura-ku, Saitama-city, Saitama 338-8570, Japan;
$^{t}$ Rikkyo University, 3-34-1, Nishi-Ikebukuro, Toshima-ku, Tokyo, 171-8501, Japan;
}	}

\abstract{We perform simulations of Cherenkov Telescope Array (CTA) observations
of a young supernova remnant RX~J1713.7--3946. This target is not only
one of the brightest sources ever discovered in very high-energy (VHE)
gamma rays but also well observed in other wavebands. In X-rays, the
emission is dominated by synchrotron radiation, which links directly
to the existence of high-energy electrons. Radio observations of CO
and H$_{\rm I}$ gas have revealed a highly inhomogeneous medium surrounding the
SNR, such as clumpy molecular clouds. Therefore gamma rays from
hadronic interactions are naturally expected. However, the spectrum in
GeV energy range measured by {\it Fermi}/LAT indicates more typical of
leptonic emission from accelerated electrons. Despite lots of
multi-wavelength information, the competing interpretations have led
to much uncertainty in the quest of unraveling the true origin of the
gamma-ray emission from RX~J1713.7--3946. CTA will achieve highest
performance ever in sensitivity, angular resolution, and energy
resolution. We estimate CTA capability to examine the emission
mechanisms of the gamma rays through simulated spatial distribution,
spectra, and their time variation.}

\FullConference{The 34th International Cosmic Ray Conference,\\
		30 July- 6 August, 2015\\
		The Hague, The Netherlands}

\begin{document}
\section{Introduction}

More than 100~years have passed since the discovery of cosmic rays but its origin has been long in question despite many observational and theoretical researches.  
The observed spectrum of cosmic rays is a power-law shape and has a break around $10^{15.5}$~eV which is so-called ``knee''.
Cosmic rays below the knee energies are thought to be accelerated somewhere in our Galaxy. 
One of the most probable candidates are supernova remnants (SNRs),
where the diffusive shock acceleration may work at the shock front of SNR blast waves.
Evidence for electron acceleration has been identified by the detection of synchrotron emission 
with spatially thin filamentary structures at shells of young SNRs (e.g., \cite{koyama95}).
On the other hand, gamma rays with hadronic origin was detected by Large Area Telescope (LAT) onboard {\it Fermi} from middle-aged SNRs IC443 and W44, which are known to be interacting with molecular clouds (MCs)\cite{ackermann13}.
The observed gamma-ray spectra are interpreted as neutral pion decay,
which is characterized by a cutoff below 300~MeV, 
due to the interaction between accelerated cosmic-ray hadrons and MCs.
However, it is also observed that that the gamma-ray spectra are suppressed above 100~GeV.
We hence expect that young SNRs could be more plausible as 
cosmic-ray accelerators to PeV energies, i.e. PeVatron.

The Cherenkov Telescope Array (CTA) is a next-generation of Imaging Air Cherenkov Telescopes (IACT) observatory
which consists of array of the large, middle, and small-sized telescopes expanding over km$^2$ area\cite{actis2011cta,ctaconcept}. 
With higher performance in comparison to the current generation IACTs, such as better spatial resolution and sensitivity, 
a search for cosmic-ray PeVatron is one of the major scientific objectives of the CTA.
A young SNR RX~J1713.7$-$3946 is one of the brightest Very High-Energy (VHE) gamma ray sources and 
spatially extended emission was observed \cite{enomoto02,aharonian04,hess06,hess07}. 
The VHE gamma-ray spectrum of RX~J1713.7--3946 is the most precisely measured 
over a wide energy band from 0.3 to 100 TeV.
Besides, plenty of multi-wavelength observations have been performed.
{\it Fermi} measured the gamma-ray spectrum of RX J1713.7--3946 in the 3--300~GeV energy range,
where the observed photon index of $1.5\pm0.1$ is favorable for inverse-Compton emission
from accelerated electrons with a spectral index of 2.0\cite{fermi1713}.
X-ray emission is dominated by synchrotron radiation
with a good spatial correlation between VHE morphology,
although the angular resolution is not good enough to be conclusive.
On the other hand, radio observations of CO and H$_{\rm I}$ gas 
have revealed a clumpy molecular clouds (MCs) surrounding the SNR,
and reported evidence for interaction between the MCs and the SNR shock (e.g. \cite{fukui03,fukui12,sano14}.
The hadronic gamma-ray emission is naturally expected
to reproduce the obseved spetrum (e.g. \cite{gab14}) .
However, our idea here is that,
if the hadronic gamma rays do exist, such component might be hidden 
by the dominant leptonic gamma-ray emission. 
It is of a great interest whether the improved sensitivity of the CTA 
could detect the possible but dim hadronic gamma rays.
Hence this object is a very good target for deep observations, and also for constraining theoretical models of cosmic-ray acceleration.


\section{Aims and methods of simulations}
The major purpose of our simulation studies is to show an example of analysis strategy
when we will obtain real data,
and to evaluate the capabilities of CTA on finding a clue for the hadronic gamma rays.

First we perform morphological analyses
in order to find out the dominant component of the VHE gamma-ray emission from RX J1713.7--3946.
As for the hadronic gamma rays, 
the morphology should be related to spatial distribution of the accelerated protons
and that of the interacting matter density which is indicated 
by the CO and H$_{\rm I}$ morphology obtained by the radio observations.
Since we currently do not know the CR distribution,
we here roughly assume that CR would be filled homogeneously inside the SNR.
On the other hand, 
the morphology of the leptonic gamma-ray emission may be traced by that of synchrotron X-ray.
We should note that the X-ray and VHE morphologies are not always completely same.
The brightness of the synchrotron emission reflects 
not only for spatial electron distribution but also local magnetic fields. 
And the inverse-Compton process is also coupled to energy densities of target photons
which includes infrared and/or optical photon field 
in addition to the cosmic microwave background.
However, the overall structure could be approximated by the X-ray morphology.
We hence apply the radio and X-ray images as templates 
for the hadronic and leptonic gamma-ray morphology, respectively.
We perform maximum likelihood test to quantitatively determine which component dominates the VHE emission from the SNR.
The templates also contain spectral information. 
Here we simply assume the same spectral shape is assumed over the SNR image.
The spectrum of the leptonic component is modeled as
\begin{equation}
\frac{ {\rm d}N_1(E)}{ {\rm d} E} = A_1  \left(\frac{E}{\rm TeV}\right)^{-\Gamma _e}\exp\left( -\frac{E}{E_{\rm c}^e}\right) ~~,
\end{equation}
where $A_1$ is a normalization factor, $\Gamma _e$ is a photon index,
and $E_{\rm c}^e$ is a cutoff energy. 
Input values for $\Gamma _e$ and $E_{\rm c}^e$ are 2.04 and 17.9~TeV, respectively, 
as reported by H.E.S.S. observations\cite{hess07}.
For the hadronic emission,
the spectrum is described as follows,
\begin{equation}
\frac{ {\rm d}N_2(E)}{ {\rm d} E} = A_2  \left(\frac{E}{\rm TeV}\right)^{-\Gamma _p}\exp\left( -\frac{E}{E_{\rm c}^p}\right) ~~,
\end{equation}
where  $A_2$ is a normalization factor, $\Gamma _p$ is a photon index,
and $E_{\rm c}^p$ is a cutoff energy. 
We adopt $\Gamma _p =2.0$ and $E_{\rm c}^p = 300$~TeV as fiducial parameters.
Therefore $A_1$ and $A_2$, or their ratio $A_2/A_1$, are the parameter to be investigated,
requiring that the sum of the integral fluxes between 1 and 10~TeV are equal to that measured by H.E.S.S..
If the hadronic gamma ray is greater, 
we could conclude that
dominant part of the hadronic component were not accelerated to the knee energies in RX J1713.7--3946.
Searching for a spectral component that extends to PeV energies, we also look for a dimmer hadronic component by spectral analysis.
The maximum likelihood fit will be performed in order to unfold spectra for each component
and evaluate statistical significance of the hadronic gamma-ray detection.
Here the spatial templates are also considered to calculate the likelihood.

We also evaluate the capability of detecting the time variation of the spectral cutoff energy, $E_{\rm max}$, with longer time scale.
The maximum energy of the CR spectrum is determined by a balance among acceleration, cooling and escape. 
Hence $E_{\rm max}$ variation, increase or decrease, depends on the SNR age and also on acceleration theories.
In the case of RX~J1713.7--3946, $\sim$10\% variation in 10--20~years may be expected,
where $E_{\rm max}$ could vary faster in the leptonic scenario\cite{ohira10}.
Since the CTA will be operational for a few tens of years,
such a long-term study in VHE energies will become possible. 
This may be a unique approach for identifying the VHE emission mechanism.

The simulation software package that we use in this study is 
{\it ctools} version 00-07-01\cite{jur13}.
We use a preliminary instrumental response function 
which corresponds to the CTA southern array located at the candidate site Aar (southern Namibia).
When we perform the simulations,
the Galactic diffuse emission and isotropic background due to gamma-like charged cosmic rays
are taken into account for background photons in the field of view.

\section{Results}

\subsection{$\gamma$-ray image}
\begin{figure}
\centering
\includegraphics[width=\linewidth]{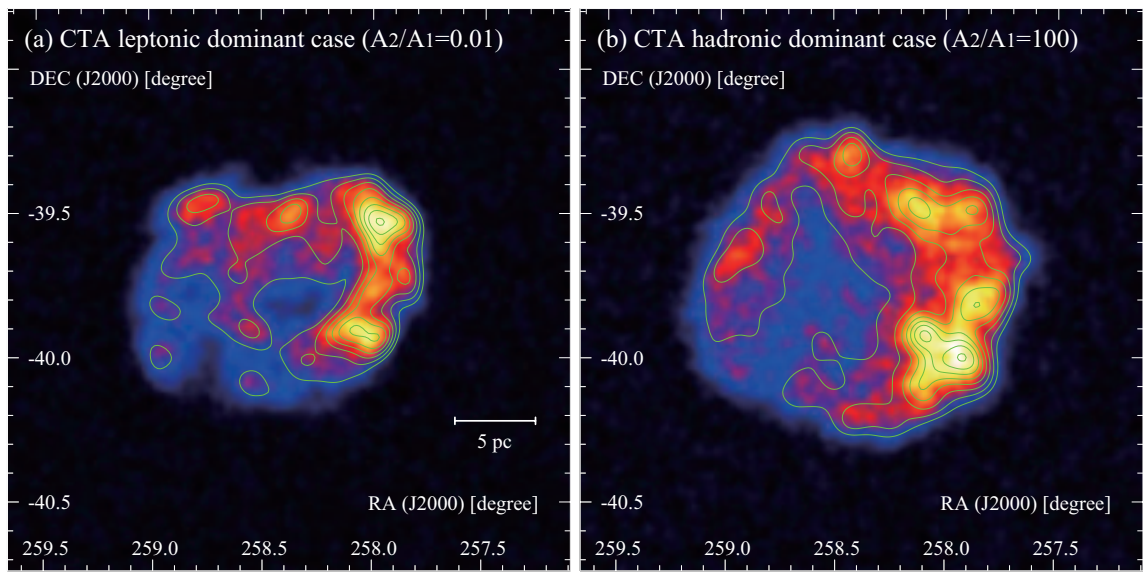}
\caption{
Simulated gamma-ray images of (a) $A_2/A_1=0.01$ (leptonic dominant case) and (b) $A_2/A_1=100$ (hadronic dominant case) with $\Gamma _p =2.0$ and $E_{\rm c}^p = 300$~TeV. The green contours show (a) $XMM$-$Newton$ X-ray intensity \cite{acero09} and (b) total interstellar proton column density \cite{fukui12}, which smoothed to match the PSF of CTA. The unit of color axis is counts pixel$^{-1}$ for both panels.
\label{fig:images}}
\end{figure}

In order to clarify the imaging capability of CTA, we first intend to simulate different gamma ray images in the energy range of 1--100 TeV by tuning $A_2/A_1$, the ratio between the hadronic and leptonic gamma rays. Figures~\ref{fig:images}a and \ref{fig:images}b show the simulated gamma-ray images in leptonic dominant case ($A_2/A_1$ = 0.01) and hadronic dominant case ($A_2/A_1$ = 100), respectively. Each gamma-ray image is similar to each overlaid contour, which corresponds to the non-thermal X rays and the total ISM protons including both molecular and atomic hydrogen, respectively. On the other hand, the spatial distributions of gamma rays are apparently different from each other, particularly the north and the southwest. In this extreme case, we can therefore determine the major component of the VHE emission by the morphological study with CTA. Incidentally, we found that $A_2/A_1$ = 1--10 showed the best spatial correspondence with the H.E.S.S. excess counts map \cite{hess07} with a correlation coefficient of $\sim$0.7--0.8. We continue to study the systematic error estimation and quantitative evaluation for the morphological difference. 

\subsection{Spectrum}
Assuming that the leptonic component is dominant,
we subsequently proceed to search for a ``hidden'' hard component with a hadronic origin.
Using reconstructed energy band of 0.5 -- 100~TeV with 50-hour observation,
likelihood analyses show the significance $>10\sigma$ to observe a dimmer hard component even for a small $A_2/A_1=0.02$.
Note that the result may be rather optimistic
since the fitting templates are the same as the input for the simulation.

\begin{figure}
\centering
\includegraphics[width=\linewidth]{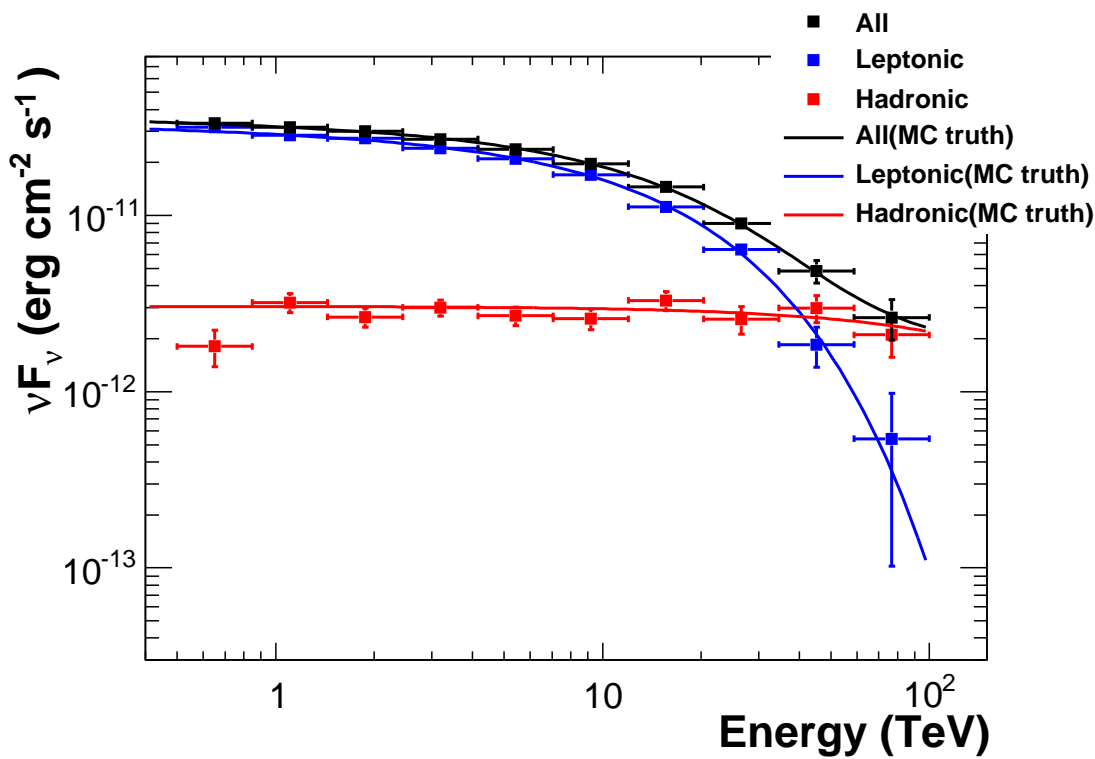}
\caption{
Spectral energy distribution of the gamma ray emission
obtained by analyzing the CTA simulation data for RX~J1713.7$-$3946 with $A_2/A_1=0.1$.
The blue and red squares are the spectral points for the leptonic and 
hadronic spatial templates, respectively.
Only statistical errors are presented.
The black squares are the total fluxes of the leptonic and 
hadronic components.
The black vertical bars are the errors for the total fluxes 
obtained by adding the errors for two components in quadrature. 
The blue, red, and black solid lines show the input spectra for the leptonic component, 
the hadronic component, and the total, respectively.
\label{fig:bin_by_bin_spec}}
\end{figure}

We then proceed to perform maximum likelihood fittings for the simulation data (with a ratio $A_2/A_1= 0.1$, for the safety) 
in 12 logarithmically spaced energy bands. 
Figure~\ref{fig:bin_by_bin_spec} shows the resulting spectrum from our `bin-by-bin' analysis
of the same 50~hr of simulation data. 
It is clear that our likelihood fits reproduce the simulated spectrum for each spatial template (i.e. hadronic or leptonic morphology),
which demonstrates the capability of detecting the hidden hadronic component in the best case scenario.

\subsection{Time variation of cutoff energy}
Detecting the time variation of the cutoff energy of the gamma ray spectrum can provide a clue to the emission scenario and acceleration theories.
The simulations are performed for three sets of intrinsic cutoff energies at 17.9 (nominal), 19.7 ($+10$\% case) and 16.1 ~TeV ($-10$\% case) each.
As a start, we consider a variation of $\Delta E_{\rm c}/E_{\rm c} = \pm 10\%$ to show how sensitive CTA will be to such fractional changes in the spectral cutoff.
Here we consider the pure leptonic scenario as an example,
and used 0.2--100~TeV photons.

\begin{figure}[]
\begin{center}
\includegraphics[width=0.8\linewidth]{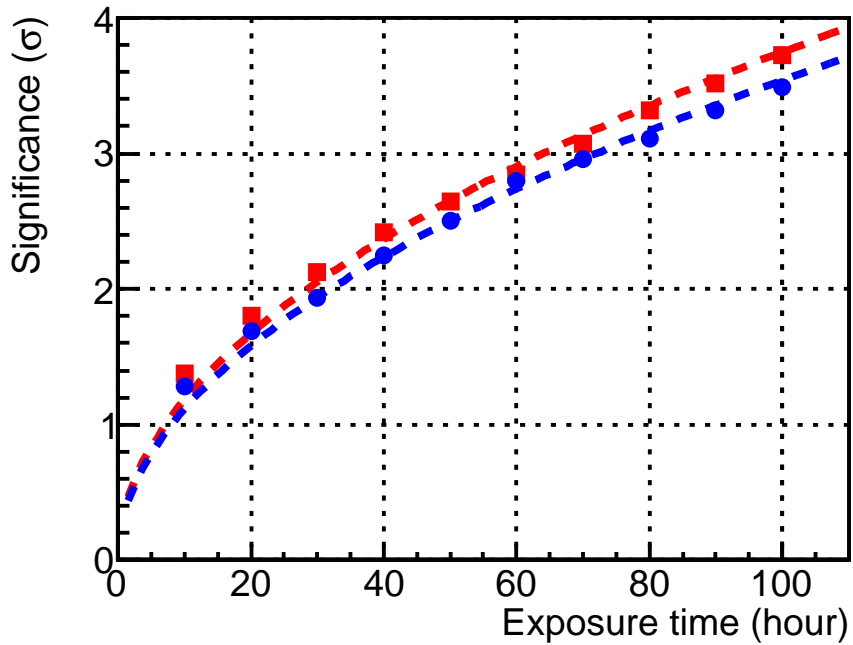}
\caption{Significance of the detected variation of $E_{\rm c}$ as a function of exposure time.
  Dashed lines represent the best-fit curve to each dataset which is proportional to $\sqrt{t}$. 
  Squares and circles represent results for the $\Delta E_{\rm c}/E{\rm c} = -10$\% 
  and $+10$\% case, respectively.
}
\label{fig-ecsigma}
\end{center}
\end{figure}

We define a significance, $s$, for the observed $E_{\rm c}$ variation as
$
  s _{\pm} (t)= \frac{|E_\pm(t) - E_0(t)|}{\sqrt{\sigma _\pm^2(t) + \sigma _0^2(t)}}, 
\label{eq:spar}
$
where $E_0$ is the nominal value and $E_\pm$ is the best fit $E_{\rm c}$ for the cases with a $\pm 10\%$ variation. 
$\sigma _0$ and $\sigma _\pm$ are the corresponding errors.
We repeat all of our simulations for 100 times and take the average of the calculated $s(t)$ for each run,
and show the result on Figure\,\ref{fig-ecsigma} .
Our result indicates that a decrease of $E_{\rm c}$ is slightly easier to identify than an increase.
As a result, a lower cutoff energy can actually be easier to measure and precisely for a given exposure.
If we observe $> 60$~hrs in the two epochs, we are able to achieve a $3\sigma$ detection for the $\Delta E_{\rm c}/E_{\rm c}=- 10\%$ case,
whereas $\sim 70$~hrs are necessary for the $+10\%$ case.
%



\section{Summary}

In this paper,
we have briefly introduced our feasibility studies for CTA observations of RX~J1713.7--3946,
mostly with 50~hrs observations.
We showed that a 50-hr observation may be enough to identify the dominant gamma ray emission component
by the morphology obtained with CTA.
And in the case that the leptonic emission would be dominant,
we should be able to quantify both the leptonic and hadronic components through spectral analysis if they are mixed with a ratio of $A_2/A_1 =0.1$ or less. 
Interestingly, we also found that CTA will be able to reveal variations of the spectral cutoff energy over 10--20 years, for the very first time.
A variation of $\Delta E_{\rm c}/E_{\rm c} = \pm 10\% $ could be detected
provided that an exposure time longer than 70~hr can be secured for the two epoch.

However we know our present study is based on a fairly simplified input model and may contain systematic errors
of which estimation is not trivial. 
And also this study can be extended to use more theoretically justified models.
More studies are left for future works.

\acknowledgments
We gratefully acknowledge support from the agencies and organizations 
listed under Funding Agencies at this website: http://www.cta-observatory.org/.
We thank S. Katsuda for providing the original \textit{XMM-Newton} image.

\end{document}